\documentclass[useAMS,usenatbib]{mn2e}
\def\ae{\hbox{AE\,Cir}}
\usepackage{graphicx}
 \usepackage{times}
\title[AE\,Circinus]{The remarkable properties of the symbiotic star AE Circinus}
\author[R.E. Mennickent, J. Greiner, J. Arenas,  G. Tovmassian, E. Mason, C. Tappert and C. Papadaki]
{R. Mennickent$^{1}$,\thanks{E-mail: rmennick@astro-udec.cl.
Based on observations carried out at ESO and CTIO telescopes: ESO proposals 65.H-0410(A) and 66.D-0141(A) and NOAO proposals 04B.0086 and 05A.0004.
}
J. Greiner $^{2}$, J. Arenas$^{1}$,  G. Tovmassian$^{3}$, E. Mason$^{4}$, C. Tappert$^{5}$, C. Papadaki$^{4,6}$\\
$^{1}$Departamento de F\'{\i}sica, Facultad de Ciencias F\'{\i}sicas y
Matem\'aticas,
Universidad de Concepci\'on, Casilla 160-C, Concepci\'on, Chile \\
$^{2}$Max-Planck-Institut f\"ur extraterrestrische Physik, 85748 Garching, Giessenbachstr 1, Germany\\
$^{3}$Instituto de Astronom\'{\i}a IA-UNAM, Apartado 70-264, 04510 Mexico D.F., Mexico\\
$^{4}$European Southern Observatory, Casilla, 19001, Santiago 19, Chile\\
$^{5}$Departamento de Astronom\'{\i}a y Astrof\'{\i}sica, Pontificia Universidad Cat\'olica, Casilla 306, Santiago, 22, Chile\\
$^{6}$Vrije Universiteit Brussel, Pleinlaan 2, 1050 Brussels, Belgium}
\begin{document}

\date{Accepted XXX. Received XXX; in original form XXX}

\pagerange{\pageref{firstpage}--\pageref{lastpage}} \pubyear{2006}

\maketitle

\label{firstpage}

\begin{abstract}

We present new optical spectroscopy and photometry,
2MASS infrared observations and 24 years of combined AAVSO and AFOEV photometry
of the symbiotic star candidate \ae.
The long-term light curve is characterized
by outbursts lasting several years and having a slow decline of
$\sim 2 \times 10^{-4}$ mag/day. 
The whole range of variability of the star in the $V$ band is about 4 magnitudes. 
The periodogram of the photometric data reveals strong signals at $\sim$ 342 and 171 days.
The presence of the emission feature at $\lambda$ 6830 \AA~ at minimum and the detection of absorption lines
of a $\sim$ K5 type star confirm the symbiotic classification 
and suggest that AE\,Cir is a new member of the small group of
s-type yellow symbiotic stars. We estimate a distance of 9.4 kpc.
Our spectrum taken at the high state shows a much flatter 
spectral energy distribution, the disappearance of the $\lambda$ 6830 \AA~  emission
feature and the weakness of the He\,II 4686 emission relative to the Balmer emission lines.
Our observations indicate the presence of emission line flickering in time scales of minutes in 2001.
The peculiar character of \ae~ is revealed in  the visibility of the secondary star at the high and low state,
the light curve resembling a dwarf nova superoutburst and 
the relatively short low states.  The data are hard to reconciliate with
standard models for symbiotic star outbursts.



\end{abstract}


\section{Introduction}

AE\,Cir (S32 = StHA32, $\alpha_{2000}$ = 14:44:52.0, $\delta_{2000}$ = -69:23:35.9)
was classified as a RCB star in the Fourth General
Catalog of Variable Stars. This classification
was rejected by Kilkenny (1989), who found
strong emission lines of H\,I and weaker emission lines of
He\,II and other species in one spectrogram spanning a
wavelength range of 3400-5100 \AA. Kilkenny noted the lack of
forbidden lines and the anomalous strong He\,II 4686 emission
for this object, and suggested a symbiotic classification, although no lines of the cool
component were detected. He also mentions the
photometric variability of the star, ranging from 12 to 14 mag. (from visual estimates)
in hundreds of days. $B,V$ photoelectric measures by Lawson \& Cottrell (1990) taken
in an interval of 107 days show the star with $V$ magnitude between 13.52 and 14.51
and color $B-V$ between 0.95 and 1.43, the object being  redder when fainter.
Based on the spectroscopy by Kilkenny, AE\,Cir was listed as a suspected symbiotic in the catalog of
Belczynski et al. (2000). Symbiotic stars have been reviewed recently by Miko{\l}ajewska (2007).
In this paper we present new photometry and low and high-resolution spectra of AE\,Cir
and investigate their emission line properties and symbiotic nature.

In Section 2 we give details of our spectroscopic observations
and show the methods used in data reduction and analysis. In that section
we also analyze available visual long-term photometric records
along with our own CCD
photometry. In Section 3 we analyze our data,
giving the main results of our research. In Section 4 we discuss our results in the context of
symbiotic stars and present our conclusions in Section 5.

\section{Observations and data reduction}

In this section we present photometric time series obtained at La Silla and
Cerro Tololo Inter American Observatory (CTIO) during several seasons. We also present
medium resolution spectra of \ae~ obtained in May 29 and 30, 2000 with the 1.54 m Danish telescope
at La Silla European Southern Observatory (ESO), high  resolution spectra obtained at the Paranal Observatory
with the ESO-VLT-UT2 telescope during March 16, 2001 and a medium resolution spectrum obtained
at the CTIO with the 4m Blanco telescope during May 04, 2007. Instrumental setups and observational
details  are given for all observing runs in this section. The spectroscopic observing log is presented
in Table 1.

\begin{table*}
\begin{minipage}{155mm}
\caption{Log of spectroscopic observations. N is the number
of spectra. We give the resolving power $R$,
HJD (-2\,450\,000) at the start and the end of the
observations and the mean seeing value.}
  \begin{tabular}{lrcrcccccc}
\hline

Date(UT)  &Observatory &     N &HJD(start) &HJD(end) &  Grism/mode &Range(\AA) &$R$ & exptime (s) & slit/seeing (")\\
\hline
29/05/00& La Silla& 1 &1694.7079 &1694.7079 &15 &3500-9500  &400  &300&1.5/1.5\\
29/05/00& La Silla& 24& 1694.7152& 1694.7724& 7 &3840-6820  &1500 &120&1.5/1.5 \\
30/05/00& La Silla& 64& 1695.6748& 1695.8160& 7 &3840-6820  &1500 &120&1.5/1.5 \\
30/05/00& La Silla& 5 &1695.8226 &1695.8305 &5  & 5080-9370 &800  &120&0.9/1.8  \\
16/03/01& Paranal &28 &1984.8623 &1984.9103 &red& 5175-6075 &50000&120&0.9/1.8  \\
16/03/01& Paranal &28 &1984.8623 &1984.9103 &red& 6119-7105 &50000&120&0.9/1.8  \\
16/03/01& Paranal &15 &1984.8224 &1984.8519 &blue& 4014-5241&50000&150&0.9/1.8  \\
04/04/07& Tololo  & 1 &4194.5472 &4194.5472 &KPGL3-1&3700-7230 &2000 &900&1.0/NA \\
\hline
\end{tabular}
\end{minipage}
\end{table*}

\subsection{ESO and CTIO photometry}

AE\,Cir was observed with the Dutch 0.9m telescope at  La Silla ESO observatory
in the night of  5/6 April 2001 in the $V$ band to check for short-term variability.
In a total of 82 images with 150 sec exposure time each, AE\,Cir is
seen at constant brightness of $V$ = 13.65 $\pm$ 0.10 with a scatter of
$\pm$ 0.03 mag, consistent with statistical fluctuations.

We also observed AE\,Cir with the CTIO 1.3m telescope operated
under the SMARTS consortium in the seasons 2004 and 2005. Observations
were done in pairs of two exposures in $B$ and $R$ each, on a total
of 39 nights. Relative photometry was obtained using daophot in IRAF\footnote{IRAF is distributed by the National Optical
Astronomy Observatories, which are operated by the Association of Universities for Research in
Astronomy, Inc., under cooperative agreement with the NSF.}.
Absolute photometric calibration was done against the USNO-B1 star
0205-0393810 with $B$ = 15.0 and $R$ = 13.9 mag, respectively.
Our relative photometry is accurate to 0.05 mag, as verified against
a number of nearby constant stars.

\subsection{ESO DFOSC low resolution spectroscopy}

The Danish Faint Object Spectrograph and Camera (DFOSC)
was used with several grisms yielding a combined
wavelength range of $\lambda \lambda$ 3500-9500 \AA. A slit width of 1.5\arcsec 
was chosen in order to match the typical point
spread function at the focal plane of the telescope.
This resulted in spectral resolutions of 4 \AA~ (grism 7),
7.5 \AA~ (grism 5) and 14 \AA~ (grism 15). The instrument
was rotated before exposures to ensure optimal orientation
relative to the parallactic angle. Reductions were
done in the usual manner with IRAF.
The spectra were flux-calibrated with the standard stars LTT\,3864
and LTT\,7987 
observed with slit width of 5\arcsec 
and were calibrated in wavelength
with He-Ne exposures taken typically after half-an-hour
sequences. A typical $rms$ of the wavelength calibration
function for grism 7 was 0.1 \AA~ (5 km s$^{-1}$ at H$\alpha$) and the
zero point for the flux calibration, corrected for the finite slit width,
is accurate at the level of 15\%.


\subsection{VLT UVES high resolution spectroscopy}

We used the high-resolution optical spectrograph of the VLT,
UVES, in visitor mode. Our setup and the slit of 0.9\arcsec 
yielded a resolving power of $\sim$ 50.000.
We obtained spectra in the blue spectral range using the blue arm
and cross disperser CD\#2 with central wavelength $\lambda$ 4630 \AA,
and red spectra using the red mode and cross disperser CD\#3 centered at $\lambda$ 6100 \AA.
The two CCDs in the red camera provided two quasi-simultaneous spectra separated
by a gap centered on the central wavelength. Reductions were made using the UVES pipeline
software, providing wavelength calibrations - based on fits of Th-Ar lines -
with an accuracy of 1.6 pixels (about 0.06 \AA~ $rms$, or 4 km s$^{-1}$ at H$\beta$).
The atmospheric absorption bands in the red spectra
were removed from the spectra using the "telluric" IRAF task. For that we used
a template obtained by normalizing the standard star spectrum to the continuum with a high order
polynomial. This high-order function was successful in fitting not only the continuum, but also
the relatively broad stellar lines of the template without affecting the narrow telluric lines.
This method allowed a good removal of telluric lines in our red spectra.
A flux calibration was also performed with the
pipeline, with the zero point accurate at the level of 10\%.
A correction factor to the photon counts was applied to the spectra to account for the
finite slit width and the seeing disc size reported in the
ESO Ambient Conditions Database.

\subsection{CTIO medium resolution spectroscopy}

We obtained one spectrum of AE\,Cir on May 04, 2007 with the RC spectrograph mounted in the CTIO
4m Blanco telescope at a resolution of 4.82 \AA.
Reductions were done in the usual manner with IRAF, calibrating in wavelength with HeNeAr lamps
that provided 62 lines that were fit with a 6th order Chebychev function resulting in a
$rms$ of 0.24 \AA~(11 km s$^{-1}$ at H$\alpha$). 
We also flux calibrated our spectrum
with the aid of the standard star LTT 4816 observed at the beginning of the night.
No telluric correction was applied for this spectrum.

\section{Results}

In this section we present and discuss photometry obtained by the
American Association of Variable Star Observers (AAVSO) and the
Association Francaise des Observateurs d'Etoiles Variables (AFOEV) that is
available at http://www.aavso.org/ and
http://cdsweb.u-strasbg.fr/afoev/, respectively.  Our own photometry
is also discussed along these data.
We also give a general description of the spectra of  our target and
discuss the published infrared photometry. We show evidence for a cool stellar component and
determine its spectral type. We also provide
an estimate of the reddening and distance and present spectroscopic measurements for the
emission lines. The emission line ratios are compared with model predictions and physical conditions of the line emitting region
are estimated.

\subsection{AAVSO, AFOEV and CTIO long-term photometry}

\begin{figure}
\scalebox{.9}[.9]{\includegraphics[angle=0,width=10cm]{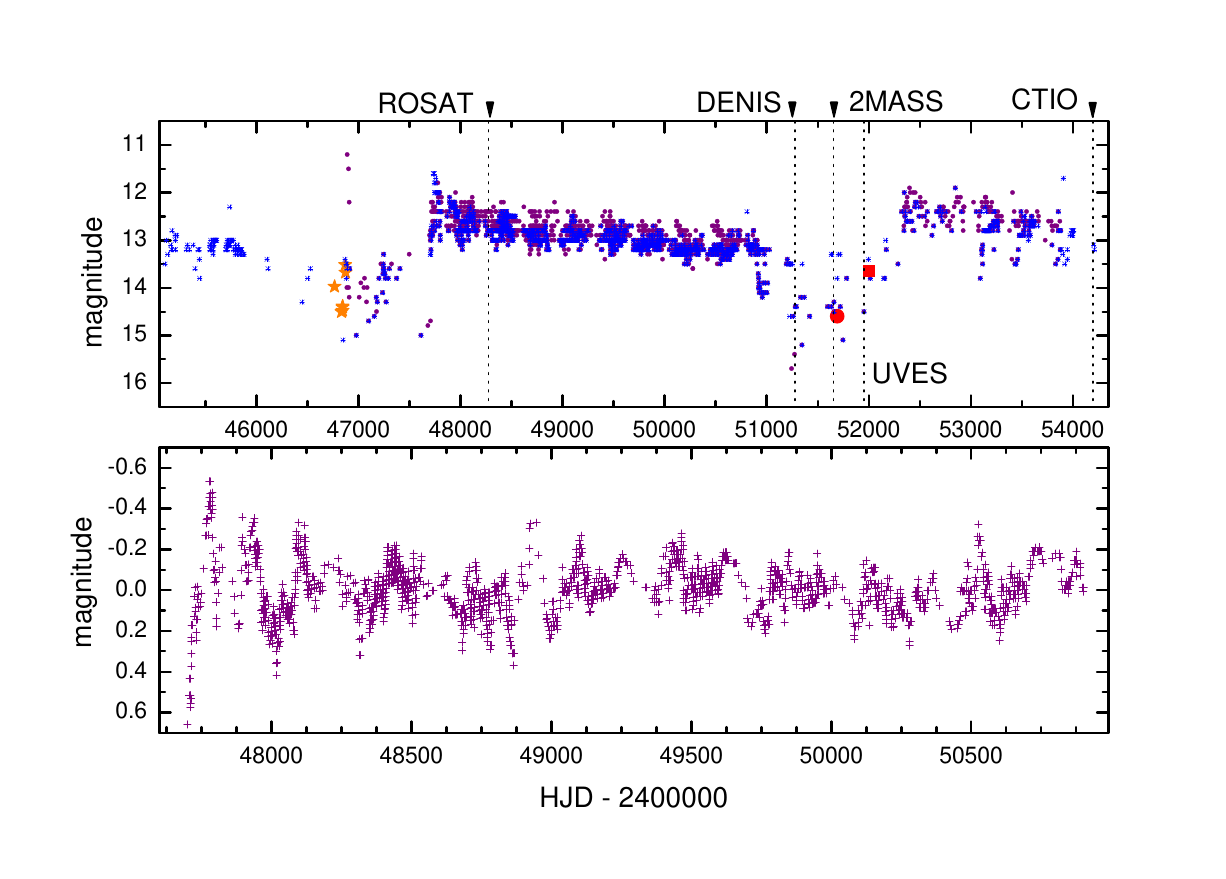}}
  \caption{Above: The AAVSO (asterisks) and AFOEV (dots) light curve of AE\,Cir along with the Lawson \& Cottrell (1990) magnitudes (stars), our $V$ magnitude obtained at La Silla in 2001 (filled square) and our
spectrophotometric $V$ magnitude of year 2000 (filled circle). We indicate times for 2MASS, DENIS, ROSAT and our 2001 and 2007 spectroscopic observations. Below: Smoothed AFOEV data after subtracting the long-term tendency showing variability in time scales of months.}
\label{spectra}
\end{figure}

24 years of visual photometry are summarized in Fig.\,1.
In spite of the limited accuracy of visual estimates (formally $\pm$ 0.1 mag, but probably $\pm$ 0.2 or 0.3 mag), the
general long-term pattern of variability appears clear. After a prolonged  epoch of slow decline from maximum,
the star rapidly enters in a faint stage, reaching magnitudes usually $\sim$ 3 mag fainter than those
observed at maximum.
The  minimum lasts $\sim$ 900 days and is characterized by irregular short-term variability.
Then the star recovers its maximum brightness and the cycle starts again.
The prominent maxima observed at HJD 2\,447\,759 and 2\,452\,390 indicate a time scale of recurrence of about 4630 days and the whole range of variability of the star in the $V$ band is about 4 magnitudes. 
However, this pattern does not repeat strictly periodically; the duration of the minimum and the subtle
characteristics of the light curve are different within the few cycles covered by the data.
We note that the star shows larger amplitude variability  during the broad
minimum than in maximum. We also note that the main outburst covered by the data could have been
started rapidly, in just 16 days. This rapid rising time is based on two datapoints, at HJD 2\,447\,614 and 
HJD 2\,447\,684. Another observation at HJD 2\,447\,708, i.e. 
already inside the outburst, also indicate a magnitude around 14.8 and if true should indicate
a short fading episode. The outburst rising time will be relevant for the interpretation in Section 4.
On the other hand the jump to the outburst with maximum around 2\,452\,390 seems to be preceded by a gradual increase of brightness. This could be the rule for AE\,Cir 
if we neglect the three mentioned  datapoints around the beginning of the main outburst.  
On the other hand, we doubt about the reality of the maxima at HJD 2\,446\,892, 2\,446\,905 and 2\,446\,913,
since they indicate $V$ about 2 mag. brighter than the average measured from 
nearby observations. Both outbursts 
show a slow decline of $\sim 2 \times 10^{-4}$ mag/day. We observe also the final part of a third
outburst at the beginning of the time series.
Our $V$ spectrophotometric magnitude of year 2000, 
the ESO $V$ average magnitude of year 2001,
and the times for our spectroscopic 
observations are also shown in Fig.\,1, indicating that we observed the star
at minimum in years 2000 and 2001 and during the high state in year 2007.

We analyzed the 5-point average smoothed AAVSO light curve during the first long decline
between HJD 2\,447\,700 and HJD 2\,450\,900. The long-term decline was subtracted
removing the best third order polynomial fit from the data.
We analyzed residuals with the pdm IRAF package (Stellingwerf 1978) finding a period of 342 $\pm$ 15 days.
A second periodicity of 171 days is also observed in the periodogram but with less power.
The  Period04 program first developed by Martin Sperl
(Lenz \& Breger 2005, http://www.astro.univie.ac.at/$\sim$dsn/dsn/Period04/), based on the Fast Fourier Transform,
yields similar results, although it favors the shorter periodicity, probably since it is more sensible to sinusoidal
variability than programs based on phase dispersion minimization algorithms.  We argue later that the
342 day periodicity provides a more realistic light curve. We note that the periods are highly significant;
the Fourier periodogram shows the short and the long periodicities at the level of 10$\sigma$ and 8$\sigma$, respectively,
where $\sigma$ is the noise amplitude (Fig.\,2). A t-test applied to 10 consecutive data bins in the phase diagram
shows inconsistency with a random distribution at the 0.05 level for all except one case.
A visual inspection of the data during minima barely reveal the presence of any periodicity. Some prominent 1-point minima observed in the light curve
do not follow any periodicity (e.g. those at JD 2\,453\,096, 2\,453\,436 and 2\,453\,835). When analyzing the data using 3 consecutive sets of 1200 days length each,
we obtain periods of 328, 364 and 364 days, respectively. Hence some period increase is possible during decline, but
the uncertainties (several tens of days) are too large for this increase to be measured.
The few times of maxima detected before the maximum around HJD\,2\,447\,700 are coherent with those of maxima after that maximum, and the amplitude decreases as the outburst evolves.
In principle the period close to 1 year could reveal a seasonal
fiducial periodicity, but the large amplitude observed in the light curve lets this appear unlikely.
The light curve phased with the 342-day period is very noisy (Fig.\,2),
probably a result of the large uncertainties of the visual estimates.
However, we can see an ellipsoidal-like variability characterized by two minima of different depth.
It is in principle possible that the 342 day period is the orbital period of a binary star.
In this case, the ephemeris for the main minimum is:  \\

$
T(\rm HJD) = 2\,447\,327.3 (\pm 3.4) + 342 (\pm 15) \times N \hfill(1) \\
$

\noindent
We note that the symbiotic nova V1329 Cyg shows a similar oscillatory behavior, with a period of
about 955 days (Chochol \& Wilson 2001). In that system, polarimetry and radial velocity
measurements prove this period to be the orbital period of the
binary system. On the other had, the symbiotic star
Z And also shows a similar behavior during decline of outburst (Formiggini \& Leibowitz 1994), but
in this case the oscillations are quasi-periodic and shorter than the orbital period.


The CTIO long-term photometric data are shown in Fig.\,3 (filled symbols), with the relevant
AFOEV data overplotted. We clearly resolve the short-term cycles,
and find that the color $B-R$ changes systematically during the
phases of minimum light. This variability is not resolved in the AFOEV data, indicating that the uncertainty
of the visual estimates is indeed larger than the formal value of $\pm$ 0.1 mag. 
We observe an eclipse-like episode around
HJD 2\,453\,620 and an additional smoother - but not completely resolved - minimum about 340 days before,
at HJD 2\,453\,280. This is about the periodicity found in the visual data.
We cannot compare these timings with the predictions of Eq.\,1, since the uncertainties involved in the period and
zero point are too large.
Note that our mean $B-R$ = 1.0 mag, as compared to that derived from
USNO-A2 of $B-R$ = 0.1 mag which is from non-simultaneous observations.
We discuss the characteristics of the light curve in Section 4.

\begin{figure}
\scalebox{.9}[.9]{\includegraphics[angle=0,width=10cm]{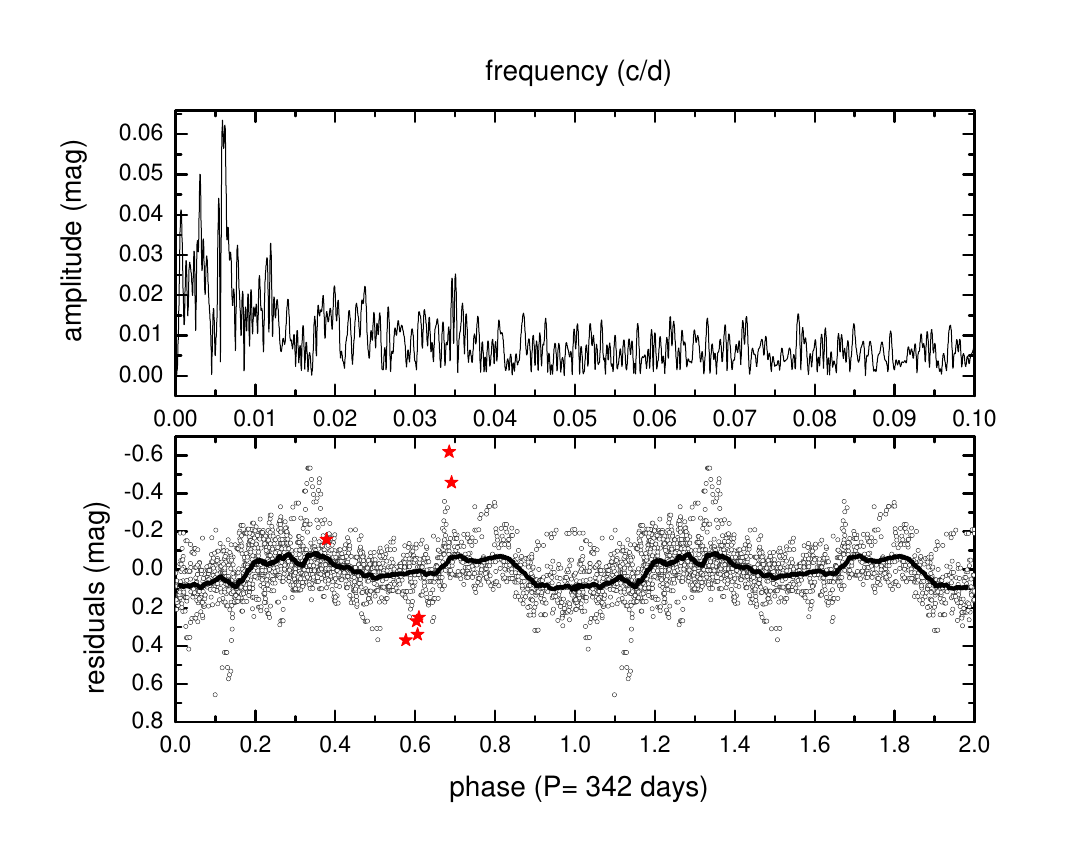}}
  \caption{Upper graph: Fourier spectrogram for the AFOEV  data taken during the long decline,
  after smoothing and removing the long-term tendency. The main peaks are around 342 and 171 days.
  Lower graph: The
  same data phased with the period 342 days and the 100-point average curve. Stars represent
  the average-subtracted Lawson \& Cottrell (1990) data.
  }
\label{spectra}
\end{figure}

\begin{figure}
\scalebox{.9}[.9]{\includegraphics[angle=-90,width=10cm]{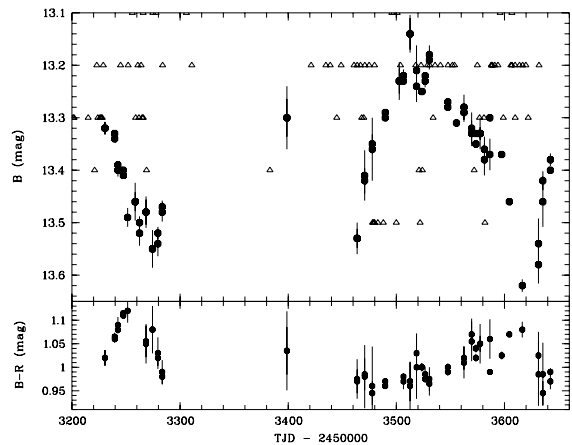}}
  \caption{Light curve obtained with the 1.3m CTIO telescope (filled symbols) along with shifted
  AFOEV data (triangles). Note the eclipse-like
  episode around HJD 2\,453\,620.}
\label{smarts}
\end{figure}

\subsection{The average spectrum and spectrophotometric magnitudes}

The averaged spectra are shown in Fig.\,4.
We observe narrow emission lines of the Balmer series and also
He\,I and He\,II emissions. The emission bump
at 6830 \AA~  typical of symbiotic stars is visible at minimum in 2000 and 2001, but not at maximum in 2007. No forbidden lines are observed.
We note high excitation lines as the very strong He\,II 4868 emission and the presence of the
C\,III-N\,III complex at 4630-4650  \AA. In 2000 and 2001 the stronger line in the complex
is N\,III 4640 \AA, but in 2007 the set of C\,III and O\,II lines around 4648 \AA is the dominant part.
The continuum slowly rises to the red in 2000 and 2001 but is quite flat in 2007. The Na\,D absorption line is strong
and visible in all epochs. We note TiO bands of
a cool star; these will be analyzed in the next section.
We also note that during 2007 the O\,II 4414 \AA~line is stronger than in previous epochs.
In 2001 we observe a larger emission strength
of He\,4686 compared to year 2000, and a reversal of the 4686/H$\alpha$ ratio.
In 2007 the object is very bright, and this coincides with a smaller emission strength for all lines, especially
He\,II 4686, that is weaker than H$\beta$. 

The magnitudes calculated from the spectra after convolution with a Johnson-V filter transmission curve are
$V$ = 14.6 in year 2000, $V$= 15.9 in 2001 and $V$= 13.04 in 2007.
These magnitudes are roughly consistent with the general pattern of the visual light curve indicating that the star
was observed at minimum during 2000-2001 and declining from maximum at 2007.
We note that our $V$ magnitude of 2001 is quite low, the faintest ever observed.
The average $V$ magnitude 13 and 20 days after our 2001 spectroscopy are 
$V$= 13.25 and $V$= 13.65, respectively, indicating  a possible jump of more than 2 magnitudes in a few days. 
However, due to the large seeing during the 2001 observations, is possible that the slit loss was larger than our correction, and therefore we consider hereafter our $V$ magnitude value as a lower limit only.
Regarding the 2007 magnitude, it fits better the tendency of the light curve in Fig. 1, but we note that although
the star was observed at the beginning of the night, when the sky was clear, 
clouds were reported 1.5 hours after finishing the AE Cir observation,
hence we give low confidence to our 2007 spectrophotometric magnitude.

\begin{figure*}
\scalebox{.9}[.9]{\includegraphics[angle=90,width=18cm]{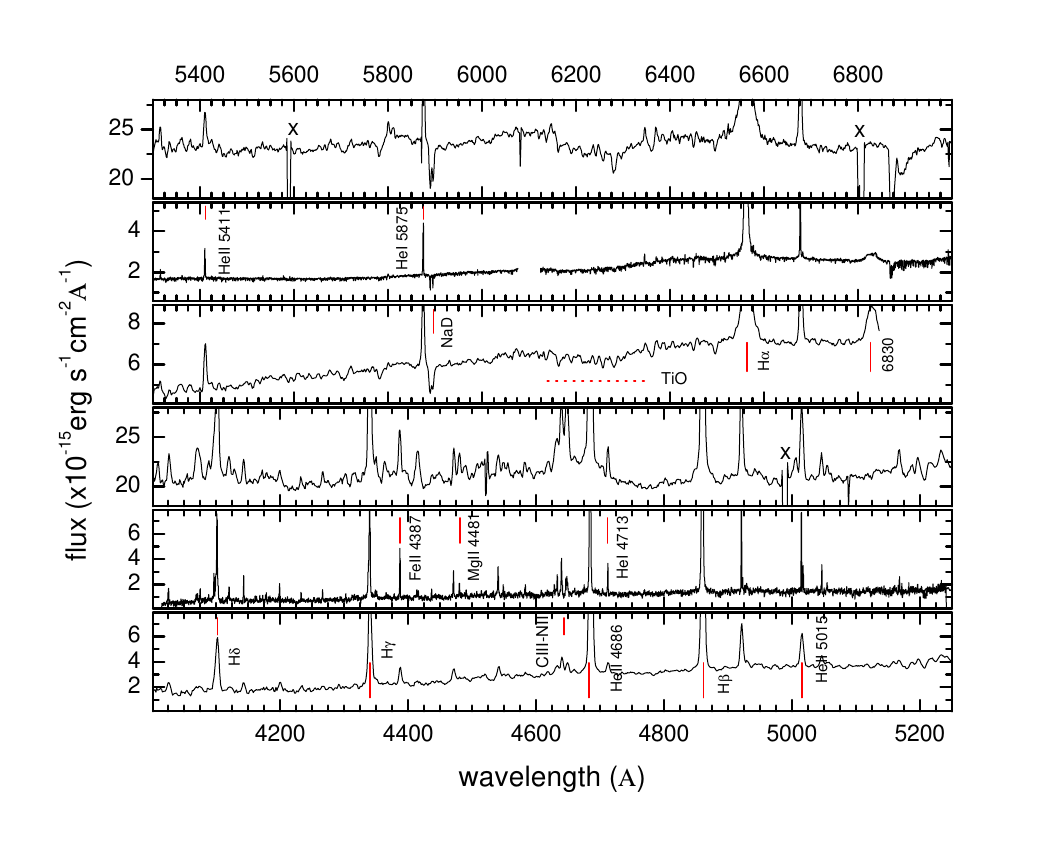}}
  \caption{The average spectrum of AE\,Cir during the observations of year 2007 (upper frames), 2001 (middle frames) and 2000 (lower frames).
  The spectra are separated in two sections and cut in the y-axis in order to increase the visualization of the spectral features. Some emission lines and the TiO band around $\lambda$ 6300 \AA~ are labeled for illustration. In the 2007 spectrum, the symbol "x" indicates bad column regions.
  }
\label{spectra}
\end{figure*}

\subsection{Detection of the cool stellar component}

Absorption lines of the cool star are clearly visible in our spectra (Fig.\,5).
We used the indices defined by Kenyon and
Fernandez-Castro (1987) to find the spectral type of the cool stellar component.

We measured these indices in the 2000 average spectra for every grating, and obtained
$[TiO]_{1}$= 0.041~ $\pm$ 0.003 and $[TiO]_{2}$= -0.004~ $\pm$ 0.009. Then, using the formula given by
Kenyon \& Fernandez-Castro (1987) we found a spectral type $\sim$ K5 consistent with both indices. The same result is
obtained with the 2007 spectrum, where we found $[TiO]_{1}$= 0.053.



In order to check if the cool star shows the same spectral type in the spectra obtained in 2001,
we subtracted  a grid of normalized UVES G8$-$M6 high resolution
stellar templates with determined MK spectral types (http://www.sc.eso.org/santiago/uvespop/)
from a region basically depleted of emission lines. The templates were allowed to vary in intensity by a scale factor and in wavelength by a
wavelength shift. The process was performed with the IRAF task "telluric", yielding the lowest $rms$ at the best stellar template compatible with the
cool star spectrum. Our grid included giants with spectral types G8, K0, K3, K5, K7, M0, M1, M4 and M6. The best solution was obtained with the
K7\,III template HD\,89736. However, the templates in the range K3 to M1 also gave a reasonable fit, differing with the K7 solution only marginally.
The method was applied in the regions 5845-6000 \AA, and 6380-6460 \AA,  obtaining consistent results in both regions (Fig.\,6).
The luminosity of the cool star cannot be established from our spectra due to contamination by line emission of the
spectral indices usually employed as luminosity indicators in late type stars (e.g. Malyuto, Oestreicher \&
Schmidt-Kaler 1997, Schmidt-Kaler \& Oestreicher 1998).


In order to determine the
velocity of the cool stellar component, we compared our high resolution spectrum in the region 5900-5970 \AA~ with a synthetic
spectrum of similar resolution calculated with the spectral synthesis
program SPECTRUM by Richard O. Gray (http://www.phys.appstate.edu/spectrum/spectrum.html, Gray \&
Corbally 1994). We run SPECTRUM with a Castelli-Kurucz atmosphere model (http://wwwuser.oat.ts.astro.it/castelli/grids.html)
with $T_{eff}$ = 3750 $K$, $\log g$ = 3.0 and solar composition.
By shifting the object spectrum in velocity until matching the spectral lines of the synthetic spectrum we calculated
a velocity, with respect to the Local Standard of Rest, of  $v_{sec}$= -25 $\pm$ 3 km s$^{-1}$ in 2001.
For the spectra of years 2000 and 2007 the resolution is
too low to obtain reliable results with this method. For these spectra we used the lines indicated in Fig.\,5 to
calculate $v_{sec}$= -4 $\pm$ 7 km s$^{-1}$ in 2000 and $v_{sec}$= +2 $\pm$ 25 km s$^{-1}$ in 2007.

 We observe that the Na\,D(1+2) equivalent width is variable: 2.50,
3.72 and 1.80 \AA~ in 2000, 2001 and 2007, respectively. The values are lower than
those expected for a K5 star, $\sim$ 5  \AA~ (Tripicchio et al. 1997).
This could mean that we never observe the cool giant completely, but veiled by some additional light in optical wavelengths.
This is more evident in the high state, where the flux is flatter and the NaD line weaker.

We investigated the infrared photometry of \ae~
extracted from the 2MASS catalog using the VizieR  on-line query form (Ochsenbein et al. 2000).
The 2MASS photometry was obtained at JD 2\,451\,624.7030, i.e. at epoch of minimum light (Fig.\,1). It
gives $J$ = 10.88 $\pm$ 0.02, $H$ = 10.05 $\pm$ 0.02 and $K$ = 9.74 $\pm$ 0.02, hence the object is much brighter in
infrared wavelengths than in the optical region at minimum. The DENIS catalog gives $I$ = 12.42  $\pm$ 0.02, $J$ = 11.01 $\pm$ 0.06 and $K$ =  9.82 $\pm$0.05 obtained at JD 2\,451\,274.7594, in general agreement with the 2MASS data.
DENIS also provides the USNOA2.0 magnitudes of $R$ = 12.7 and $B$ = 12.8 obtained at different epoch.
The 2MASS infrared colors $J-H$ = 0.83 $\pm$ 0.03 and $H-K$ = 0.31 $\pm$ 0.03 are typical for S-type symbiotics and probably exclude
the presence of dust emission at this epoch (Allen 1982). The colors of S-type symbiotics have been interpreted as
due to the presence of a late star, typically an M giant, as opposed to D-type symbiotics which contain heavily dust-reddened
Mira variables.

We do not have a determination of the color excess at
the epoch of the infrared observations. However,
assuming that AE\,Cir is beyond the galactic dust disc, it
should has at least $E(B-V)$ = 0.26, the value obtained from the  galactic extinction map by Schlegel, Finkbeiner \&
Davis (1998) in the direction of AE\,Cir ($l_{II}$ = 312.67, $b_{II}$ = -8.69).
This means dereddened colors of $(J-H)_{0}$ $<$ 0.76 $\pm$ 0.03
and $(H-K)_{0}$ $<$ 0.27 $\pm$ 0.03. The first color is consistent at the low limit with the value
expected for a K5 giant ($(J-H)$ = 0.73) but the second color is a bit redder compared with the expected
$(H-K)$ = 0.18. The above reference colors are from Koornneef (1983), after transforming to the 2MASS system using 
the equations by Carpenter (2001).

An estimate for the distance can be obtained assuming maximum visibility of the red giant at the 
minimum of $V \sim 15.5$ and using reddening $E(B-V)$ = 0.26. We obtain $d$ = 9.4 kpc assuming $M_{V}$ = -0.2 for a K5 giant. 

\begin{figure}
\scalebox{.9}[.9]{\includegraphics[angle=0,width=10cm]{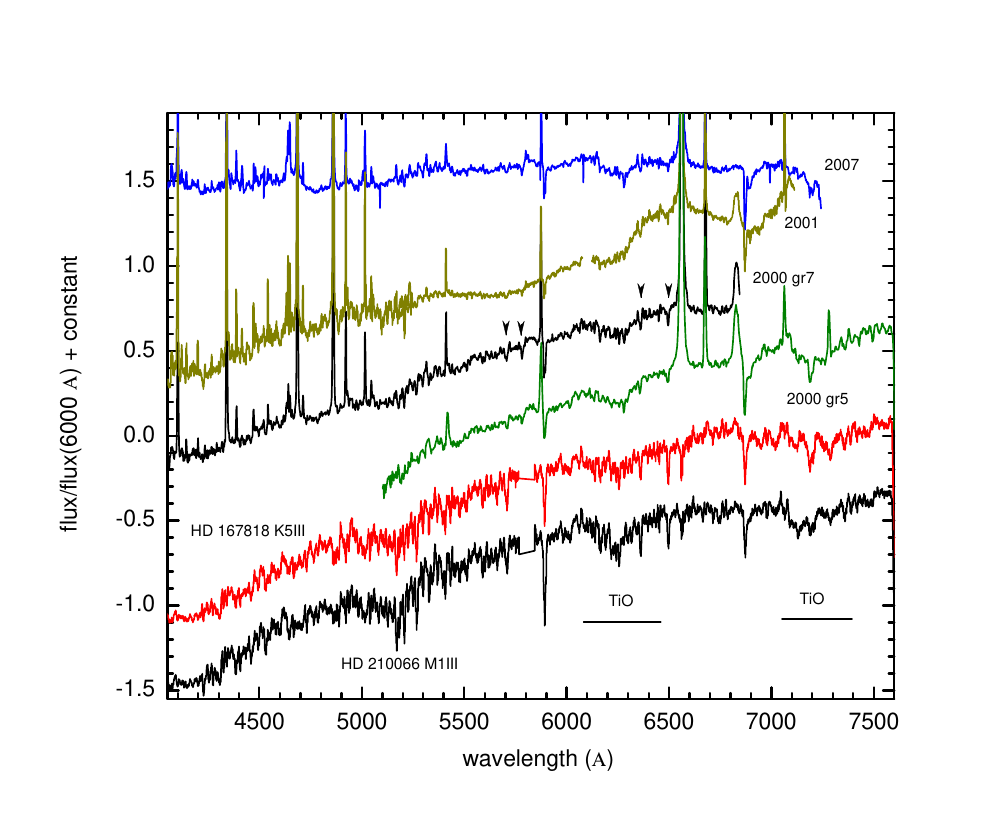}}
  \caption{The normalized spectra showing secondary star features. The TiO absorption bands used in the      determination of the
  spectral type are labeled. Arrows indicate lines used for measuring the cool star radial velocity.
  Spectra of two giants of known spectral type taken from the UVES database (see text) are shown for comparison.
  No telluric correction has been applied for the science spectra, and UVES spectra were resolution degraded
  for better comparison.
  Bad column regions of the 2007 spectrum were continuum interpolated.}
\label{fluxcomp}
\end{figure}

\begin{figure}
\scalebox{.9}[.9]{\includegraphics[angle=0,width=10cm]{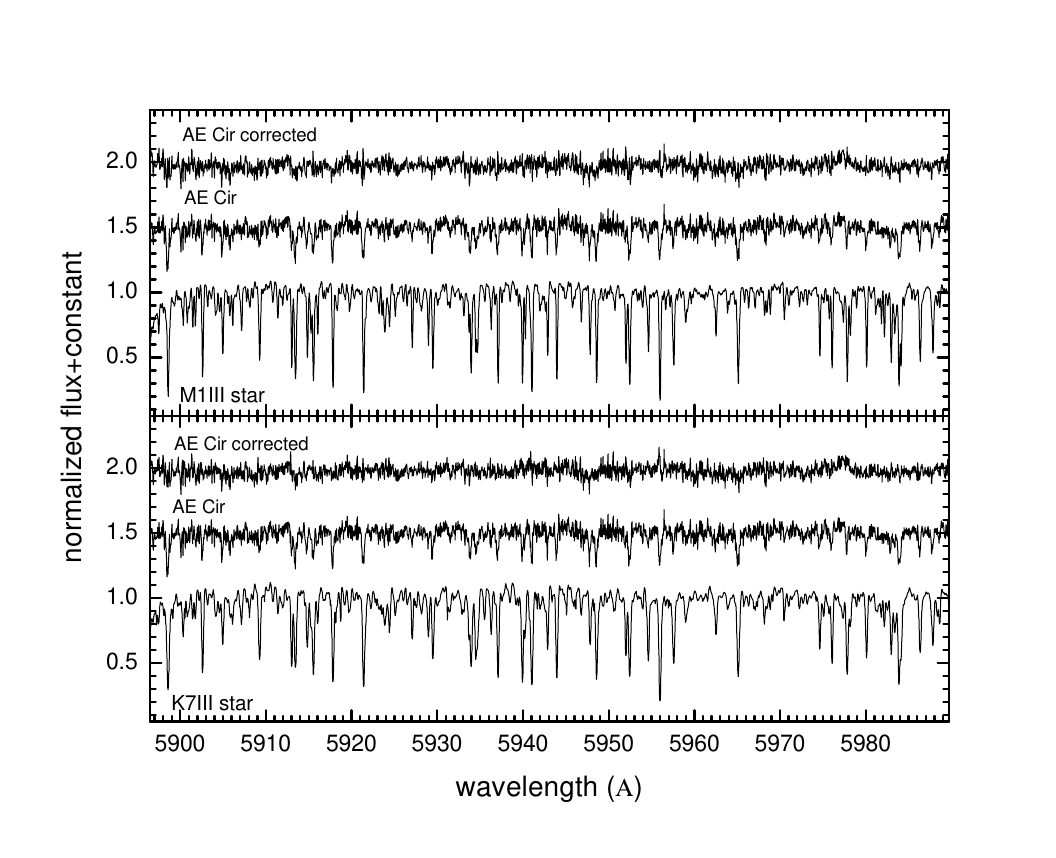}}
  \caption{Illustration of the process of searching for the spectral type of the secondary star in spectra of March 2001.
  We show the spectrum of AE\,Cir in a region depleted of emission lines, along with the spectra of M1\,III and K7\,III template stars,
  and the result of the optimal subtraction of absorption features.}
\label{}
\end{figure}

\begin{table*}
\begin{minipage}{190mm}
\caption{The emission lines observed in the spectra of \ae. We give the equivalent width $EW$,
the full width at half maximum $GFWHM$, the radial velocity at the line center with respect to the Local Standard of Rest,
the ionization potential $\chi_{\rm ion}$ and the
maximum line intensity normalized to the continuum.
The equivalent width of the Na\,D1 and Na\,D2 lines is the total equivalent width of the three components.
All parameters refer to 2001 spectra, unless specified. "nr" means line
not resolved, "a" asymmetrical line and "b" blended line.}
  \begin{tabular}{lrcrccccccc}
\hline
ion  &$\lambda_{lab}$ &$\lambda_{obs}$& $EW_{2007}$ (\AA) & $EW_{2001}$ (\AA) & $EW_{2000}$ (\AA) &  $GFWHM$ (km s$^{-1}$)& $v$ (km s$^{-1}$)& $\chi_{ion}$ (eV)& $I/I_{c}$ &note \\
\hline
H\,I      &3889.051&    -      & -   &2.7   & 11 &    -     &    -    &  -      & -    &  -              \\
H\,I      &3970.074&    -      & -   &2.5   & 14 &    -     &    -    &  -      &  -   & -               \\
unknown &-       &    4096.16  &- & 0.86   & -  &   43  &    -    &  35.1   & 3.0  &   -             \\
N\,III    &4097.310&    4096.62&- & 0.67   & -  &   20  &-37    &  47.4   & 4.4  &  -              \\
He\,II &4100.040       &  4099.43&- & 2.27   & -  &   87  &    -32    &  54.4      & 3.2  &see text                \\
H\,I      &4101.737&    4101.06&4.6 & 8.16   & 16 &   87  &   -33 &  13.5   & 9.6  &  -              \\
N\,III    &4103.370&    4102.55&- & 1.75   & -  &   87  &   -47 &  47.4   & 2.5  & -               \\
He\,I     &4120.993&    4119.96&- & 1.21   & -  &   56  &   -62 &  24.6   & 3.4  &  -              \\
He\,I     &4143.790&    4142.89&0.6 & 2.19   & 2  &   54  &   -52 &  24.6   & 4.2  &  -              \\
He\,II    &4199.830&    4198.95&- & 2.19   & 1  &   81  &   -50 &  54.4   & 3.2  &  -              \\
C\,II     &4267.020&    4266.16&0.4& 1.21   & -  &   60  &   -53 &  24.4   & 2.5  &  nr              \\
C\,II     &4267.270&    -      &- &  -     & -  &   -   &     - &   -     & -    &  nr              \\
He\,II    &4338.670&    4337.79&- & 4.56   & -  &  101  &   -48 &  54.4   & 4.4  & see text         \\
H\,I      &4340.468&    4339.77&6.6 & 17.14  & 21 &  101  &   -33 &  13.5   & 12.5 &abHeII-N\,III?  \\
Fe\,II    &4386.570&    4386.20&1.1& 1.62   & -  &   70  &   -12 &  16.2   & -    & -               \\
He\,I     &4387.928&    4387.19&- & 1.97   & 3  &   25  &   -35 &  24.6   & 6.7  & -               \\
O\,II     &4414.370&    4413.77&1.5& 0.88   & -  &   85  &   -46 &  35.1   & 1.8  &  nr              \\
O\,II     &4414.909&    -      &- &  -     & -  &   -      &     -   &  35.1   & -    & nr               \\
O\,II     &4416.975&    4415.87&- & 0.61   & -  &   69  &   -62 &  35.1   & 1.9  &    -            \\
He\,I     &4437.549&    4436.73&- & 0.61   & -  &   36  &   -42 &  24.6   & 2.2  &    -            \\
He\,I     &4469.920&    4470.76&0.7 & 2.24   & 2  &   70  &   -5 &  24.6   & 3.4  &      nr          \\
He\,I     &4471.477&       -   &- &    -   & -  &   -      &     -   &  24.6   &  -   & nr               \\
He\,I     &4471.688&       -   &- &    -   & -  &   -      &     -   &  24.6   &  -   & nr               \\
Mg\,II    &4481.129&    4480.14&0.7& 0.90   & -  &   54  &   -60 &  15     & 2.3  &       nr         \\
Mg\,II    &4481.327&    -      &- & -      & -  &   -      &    -    &  15     &  -   &  nr              \\
He\,II    &4541.590&    4540.64&- & 2.51   & 1  &   78  &   -49 &  54.4   & 3.4  &b               \\
Fe\,II    &4549.214&    4548.74&- & 0.46   & -  &   27  &   -26 &  16.2   & 2.3  &        nr        \\
Fe\,II    &4549.467&       -   &- &    -   & -  &    -  &   -     &  16.2   &  -   &      nr          \\
Fe\,II    &4583.829&    4583.11&- & 0.46   & -  &   18  &   -39 &  16.2   & 2.6  &      nr          \\
Fe\,II    &4583.990&       -   &- &  -    & -  &   -   &    -    &  16.2   &  -   &    nr            \\
Fe\,II    &4629.336&    4628.62&- & 0.40   & -  &   23  &   -33 &  16.2   & 2.2  &    -            \\
N\,III    &4634.160&    4633.28&- & 1.10   & -  &   54  &   -44 &  47.4   & 2.2  &    -            \\
N\,III    &4640.640&    4639.69&- & 2.08   & -  &   54  &   -48 &  47.4   & 3.6  &    -            \\
N\,III    &4641.900&    4640.83&- & 0.69   & -  &   54  &   -56 &  47.4   & 2.0  &bOII4642?  \\
C\,III    &4647.400&    4646.44&- & 0.71   & -  &   37  &   -49 &  47.9   & 2.6  &    -           \\
O\,II     &4649.139&    4647.69&- & 0.55   & -  &   37  &   -80 &  35.1   & 2.1  &    -           \\
C\,III    &4650.160&    4648.33&- & 0.68   & -  &   37  &   -105&  47.9   & 2.4  &    -           \\
O\,II     &4650.848&    4649.30&- &0.48   & -  &   37  &   -87 &  35.1   & 1.9  &    -           \\
C\,III    &4651.350&    4650.00&- &0.19   & -  &   37  &   -74 &  47.9   & 1.3  &    -           \\
He\,II    &4685.682&    4684.83&15.0 &53.10  & 46 &   64  &   -41 &  54.4   & 45.8 &a              \\
He\,I     &4713.143&    4712.30&0.6&1.95   & 1  &   54  &   -48 &  24.6   & 3.3  &      nr         \\
He\,I     &4713.373&       -   &- &   -   & -  &   -      &    -    &  24.6   &  -   &  nr             \\
He\,II    &4859.323&    4858.51&- &6.51   & -  &   106 &   -37 &  54.4   & 5.7  &bNIII?, see text        \\
H\,I      &4861.332&    4860.64&16.7 &22.60  & 37 &   106 &   -29 &  13.5   & 15.8 &b              \\
unknown &     -   &    4920.60&- &0.92   & -  &   42  &   22  &         & -    &    -           \\
He\,I     &4921.929&    4921.08&2.5& 3.78   & 5  &   30  &   -39 &  24.6   & 8.6  &   -            \\
Fe\,II    &4923.921&    4923.10&- &0.47   & -  &   29  &   -37 &  16.2   & 2.2  &   -            \\
Fe\,I?    &5014.950&    5014.19&- &0.76   & -  &   36  &   -32 &  7.9    & -    &bFeI?          \\
He\,I     &5015.675&    5014.82&2.1&3.19   & 4  &   37  &   -38 &  24.6   & 5.6  &   -            \\
Fe\,II    &5018.464&    5017.61&- &0.96   & -  &   27  &   -38 &  16.2   & 2.9  &ab?         \\
He\,I     &5047.736&    5046.87&0.6 &1.17   & 2  &   25  &   -38 &  24.6   & 2.8  &ab?         \\
MgI?    &5183.604&    5182.79&- &0.35   & -  &   24  &   -34 &  7.6    & 1.7  &     -          \\
Fe\,II    &5197.569&    5196.78&- &0.22   & -  &   24  &   -32 &  16.2   & 1.7  &    -           \\
Fe\,II    &5234.620&    5233.82&- &0.18   & -  &   15  &   -33 &  16.2   & 1.6  &    -           \\
Fe\,II    &5275.994&    5275.22&- &0.25   & -  &   26  &   -31 &  16.2   & 1.5  &a?             \\
Fe\,II    &5284.092&    5283.29&- &0.14   & -  &   12  &   -32 &  16.2   & 1.3  &    -           \\
unknown &-       &    5315.18&- &0.25   & -  &   53  &    -    &  -      & 1.2  &    -           \\
Fe\,II    &5316.609&    5315.80&- &0.39   & -  &   21  &   -32 &  16.2   & 1.9  &  nr             \\
Fe\,II    &5316.777&       -   &-  &  -   & -  &   -      &    -    &  16.2   &  -   & nr              \\
Fe\,II    &5362.864&    5362.03&- &0.11   & -  &   17  &   -33 &  16.2   & 1.4  &      -         \\
 \hline
\end{tabular}
\end{minipage}
\end{table*}

 \begin{table*}
\begin{minipage}{190mm}
\caption{Table 2 - Continuation.}
  \begin{tabular}{lrcrccccccc}
\hline
ion  &$\lambda_{lab}$ &$\lambda_{obs}$& $EW_{2007}$ (\AA) & $EW_{2001}$ (\AA) & $EW_{2000}$ (\AA) &  $GFWHM$ (km s$^{-1}$)& $v$ (km s$^{-1}$)& $\chi_{ion}$ (eV)& $I/I_{c}$ &note \\
\hline
He\,II    &5411.520&    5410.37&1.0& 2.95   & -  &   82  &   -51 &  54.4   & 3.0  &  -             \\
He\,I     &5875.618&    5873.53&2.65& 0.36   & 4  &   36  &   -93&  24.6   & 1.6  &bFeI?          \\
He\,I     &5875.650&    5874.35&-& 1.64   & -  &   37  &   -52 &  24.6   & 3.3  &  -             \\
He\,I     &5875.989&    5875.16&-& 2.24   & 5   &   37  &   -29 &  24.6   & 4.1  &  -             \\
NaD1    &5889.953&    5888.98&1.0 &2.50   &1.3   &   -      &   -39 &  5.1    &  -   &-              \\
-       &5889.953&    5889.32&- &-      & -  &   -      &   -19 &  5.1    &  -   &-               \\
-       &5889.953&    5889.70&- &-      & -  &   -      &   0  &  5.1    &  -   &-               \\
NaD2    &5895.923&    5894.92&0.8& 1.72   &1.2   &   -      &   -38 &  5.1    &  -   &-               \\
-       &5895.923&    5895.28&- &-      & -  &   -      &   -19 &  5.1    &  -   &-               \\
-       &5895.923&    5895.67&- &-      & -  &   -      &   0  &  5.1    &  -   &-               \\
unknown       &-       &    6345.00&-& 0.15   & -  &   32  &   -     &  -      & 1.2  &-               \\
Mg\,II    &6346.670&    6345.79&-& 0.25   & -  &   32  &   -23 &  15     & 1.4  &   -             \\
He\,II    &6560.099&    6558.73&-& 14.10  & -  &   142 &   -49 &  54.4   & 6.2  &bFeII?, see text         \\
H\,I      &6562.817&    6561.65&33.9& 60.30  & 74 &   130 &   -46 &  13.5   &14.5  &bHII            \\
C\,II     &6578.030&    6576.45&-& 0.18   & -  &   -      &   -59 &  24.4   & 1.3  &double          \\
C\,II     &6582.850&    6581.25&-& 0.12   & -  &   -      &   -60 &  24.4   & 1.2  &double          \\
Fe\,II    &6677.330&    6675.95&-& 1.24   &    &   60  &   -49 &  16.2   & 1.8  &  -              \\
He\,I     &6678.149&    6676.81&4.3& 3.96   & 6  &   40  &   -47 &  24.6   & 5.8  &abFeII?         \\
He\,I     &7065.188&    7063.77&2.4& 1.57   & 2  &   45  &   -47 &  24.6   & 2.5  &   -             \\
He\,I     &7065.719&    7064.70&-& 1.71   & -  &   35  &   -30 &  24.6   & 3.1  &   -             \\
He\,I     &7281.349&    -      &-&  -     & 2  &   -      &    -    &  -      & -    &   -             \\

\hline
\end{tabular}
\end{minipage}
\end{table*}

\normalsize

\subsection{The high resolution spectrum of year 2001}

We searched for emission lines in the averaged spectra and fit them with single
Gaussian functions or, in the case of blended lines,
with multiple Gaussian functions with variable central wavelength $\lambda_{c}$ and
full width at half maximum $GFWHM$. Some lines were not
separated at our instrumental resolution, for instance C\,II 4267.02 and C\,II 4267.27. In these cases
a single gaussian fit was performed and then the average rest wavelength used as the central wavelength
for the radial velocity calculation.
The emission bump at $\lambda$ 6830 \AA, usually observed in symbiotic stars,
is difficult to identify with a single line, their
$GFWHM$ in 2001 was 26 \AA, and its $EW$ was 8 \AA~in 2000, 5.5 \AA~in 2001 and zero in 2007.
We find no trace of the $\lambda$ 7082 \AA~ bump sometimes observed in symbiotic stars.
From the parameter $\lambda_{c}$ we measured the radial velocity
for every line with respect to the Local Standard of Rest (LSR).
Results of our analysis are shown in Table 2. We note that besides H\,I, He\,I and He\,II
emission lines, we find emission of Mg\,II, Fe\,II, O\,II, N\,III, C\,II and C\,III.
The C\,II lines detected at $\lambda$ 6576.45 \AA~ and  $\lambda$ 6581.25 \AA~ are doubles and
show a half peak separation of 33 $\pm$ 5 and 26 $\pm$ 5 km s$^{-1}$ respectively.
The lines with maximum ionization potential observed in the spectra correspond to those of
He\,II (54,4 eV). For the 2007 spectrum, the average radial velocity for 8 emission
lines is $v_{LSR}$= -44 $\pm$ 8 km s$^{-1}$.

\begin{figure*}
\scalebox{.9}[.9]{\includegraphics[angle=0,width=17cm]{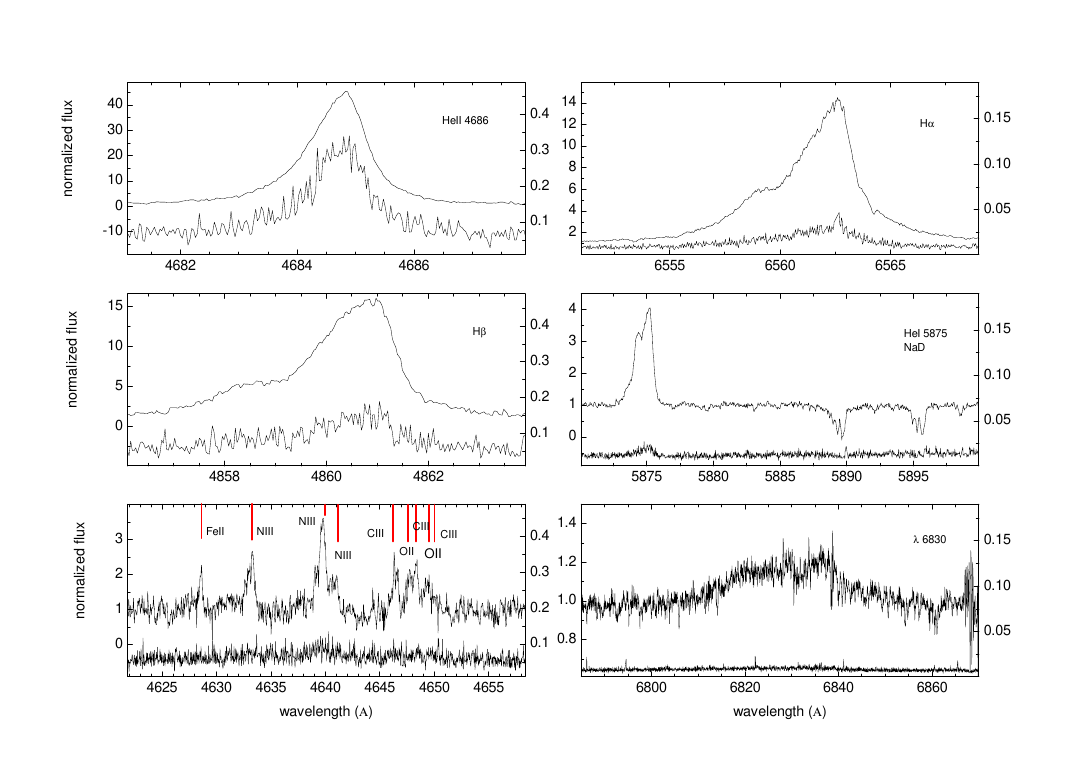}}
  \caption{Average normalized emission lines (upper profiles, labels at the left axis)
  and standard deviation (lower profiles, labels at the right axis) during March 2001. Lines of the C\,III-N\,III complex
  are labeled for illustration. Note that variability is higher in some lines, and
  mainly restricted to the line center.}
\label{spectra2}
\end{figure*}

We show in Fig.\,7 the main averaged emission lines during 2001
along with the standard deviation obtained after the
averaging process. We have averaged continuum normalized  spectra.
In Fig.\,7 we note the asymmetry
of He\,II\,4686 and Balmer lines - their maxima are displaced to the red.
The Balmer lines are very interesting. We note that the asymmetry could be due to blending with He\,II lines.
However, equivalent widths for isolated He\,II lines are of the order of 2 or 3 \AA~
whereas those obtained for He\,II lines, if they are assumed to be blended with symmetric Balmer
lines, are larger (Table 2).
We also note that in spite of the changes in emission line strength,
the ratios between the red and violet peak of the Balmer lines are very similar
(2.7, 2.8, 2.6 and 2.3 for H$\delta$, H$\gamma$, H$\beta$ and H$\alpha$ lines, respectively) and that this tendency
cannot be explained by blending with He\,II lines. Here we have interpreted the shoulders observed in the violet part of the Balmer lines as
the "violet peak" whereas the main peak is the "red peak".
We conclude that the Balmer lines are intrinsically
asymmetric and probably have a very small blending with He\,II lines.  Therefore
the parameters for the He\,II lines located near the Balmer lines
should be taken with caution. Consequently, only isolated He\,II lines are considered in our subsequent analysis.
When using the maximum intensity as
the reference for the radial velocity of the Balmer lines, we obtain values that are
-10, -19, -20, and 0 km s$^{-1}$ different from the values listed in Table 2,
for H$\delta$, H$\gamma$, H$\beta$ and H$\alpha$.
On the other hand, when using the line barycenter, we obtained
values differing by +3, +3, +12 and +43 km s$^{-1}$ from the values listed in Table 2 for the same lines.
This exercise illustrates the influence of the Balmer line asymmetry on the radial velocity measurements.
We note that the peak separation is about 150 km s$^{-1}$ for all Balmer lines. In addition,  the
H$\alpha$ emission wings extends beyond 500 km s$^{-1}$ from the velocity associated with maximum intensity,
and shows several structures. The red wing  has a steep slope and reveals
a small but significant absorption feature at $\lambda$ 6564.24 \AA. On the other hand, the blue wing has a much less
steep slope, and shows a broad bump redwards $\lambda$ 6560 \AA. The $FWHM$ measured with the cursor is 2.86 \AA~ (130 km s$^{-1}$).
This is significantly lower than the values obtained with the averaged grating-7 spectra of 2000, namely 365 km s$^{-1}$ and the
value obtained in 2007, viz.\, 552 km s$^{-1}$.
Even considering instrumental broadening the H$\alpha$ line is broader in year 2007. Contrary to other
epochs, the 2007 H$\alpha$ emission profile looks  symmetrical and double peaked with full peak separation
of 332 km s$^{-1}$. We note that when compared with our spectra, the (non flux calibrated) spectrum shown by Kilkenny (1989) reveals larger emission:
this author reports equivalent widths of -81 and -55 \AA~ for the emission lines He\,II\,4686 and H$\beta$, respectively.

The $rms$ curves in Fig.\,7 reflect the net line variability if
the continuum remained constant during our short observations.
Variability is not observed in the Na\,D lines, neither in the bump at $\lambda$ 6830 \AA. However,
we do observe variability in others lines. In general, this variability is larger in
the line center. Maximum variability is observed in the He\,II\,4686 line, amounting to 30\%  of the
line intensity. H$\beta$ is variable at the 15\% level and H$\alpha$ at the 2\% level.

During the 2001 observing session we do not observe radial velocity changes in any of the lines.
The He\,II\,4686 line, for instance, remains at the same position during 45 minutes
with a mean of 4684.697 $\pm$ 0.003 \AA~ ($rms$, 15 spectra). For this reason, we concentrated
our analysis on the average spectrum, and measured the radial velocity of their lines.
The radial velocity distribution of the lines have a broad maximum between -63 and -23
km s$^{-1}$. The average is  -42 $\pm$ 19 km s$^{-1}$ ($rms$, compare with -26 $\pm$ 20 km s$^{-1}$ in 2000 and
-44 $\pm$ 8 km s$^{-1}$ in 2007).
The radial velocity shows a slight decrease at lines of high ionization potential.
A linear least squares fit gives $v$ (km s$^{-1}$) = -30.9($\pm$6.7) - 0.59($\pm$0.22) $\chi$,
with a correlation coefficient of -0.68, where $\chi$ is given in eV.
The $GFWHM$ shows the contrary effect; it increases with the ionization potential.
A linear least squares fit, excluding the H\,I outlier, gives $GFWHM$ (km s$^{-1}$) =
27.7($\pm$7.7) + 0.64($\pm$0.24) $\chi$, with a correlation coefficient
of 0.69. In this case the H\,I lines are clearly broader and deviate from the general tendency,
a behavior already observed in Fig.\,7.
Since we observe this tendency in all Balmer lines, it cannot be due to the phenomenon of
Raman scattering, that has been suggested to explain the large
H$\alpha$ wings observed in some symbiotic stars (Nussbaumer, Schmid \& Vogel 1989).


\subsection{The $\lambda$ 6830 \AA~ emission bump}

The presence of the $\lambda$ 6830 \AA~ emission feature in AE\,Cir  is a strong argument in favor of the symbiotic classification.
At present, this feature has not been observed in any other kind of  astrophysical object (Allen 1980).
Schmid (1989) proposed that
the $\lambda$ 6830 \AA~ emission observed in symbiotic stellar systems
is the product of resonant Raman scattering
of the OVI resonance lines $\lambda$ 1032 \AA~and $\lambda$ 1037 \AA~by hydrogen, near the
Lyman $\beta$ frequency. The OVI photons are produced
in the region near the hot component  and then Raman scattered in the neutral region
near the red giant. The very existence of this line suggest a temperature
of $>$ 100\,000 $K$ for the ionizing source (M\"{u}rset \& Nussbaumer 1994) and
a scattering region with $N_{H\,I} \geq 10^{22}$ cm$^{-2}$.
The $\lambda$ 6830 \AA~ line appears as a broad emission in the 2001 spectra of AE\,Cir with two peaks on the red side at
$\lambda$ 6829.2 \AA~ and $\lambda$  6836.6 \AA~ (Fig.\,7). The peaks surround a central depression at $\lambda$
6831.8 \AA. Some narrow absorptions are also observed in the profile, and they could be 
part of the spectrum of the cool star.
A single Gaussian fit to the overall $\lambda$ 6830 \AA~ emission line
gives $EW$ = 4.66 \AA, $GFWHM$ = 23.3 \AA, and $\lambda_{c}$ = 6829.32 \AA.
The  emission line central wavelength compares well with the mean value of $\lambda_{c}$ = 6829.5 $\pm$ 0.5 \AA~ for 60 symbiotic
stars given by Allen (1980).
The shape of the line suggests the presence of many
components, and the emission wings are extended, as in  the case of H$\alpha$,
but reaching velocities up to $\sim$ 1000 km s$^{-1}$.

\subsection{The $\lambda \lambda$ 4640-4650 emission and possible X-ray excitation}

Emission around $\lambda$ 4640 \AA~ has been detected in X-ray sources and also in
non X-ray emitters like novae, Wolf-Rayet stars, planetary nebulae, gaseous nebulae,
symbiotic stars, Of, Be, and P Cygni stars. McClintock, Canizares and Tarter (1975)
suggested that this emission comes from ions of C\,III, N\,III and O\,II. They also
discussed three possible origins for this emission; non-selective emission by one or more of the ions
C\,III, N\,III and O\,II,
selective emission of N\,III $\lambda \lambda$ 4634-4641 in the "Of star process" and
selective emission of  N\,III $\lambda \lambda$ 4634-4641 in the Bowen fluorescence process.
They argue that it appears likely that the Bowen fluorescence mechanism is an important
source for N\,III  $\lambda \lambda$ 4634-4641 emission in X-ray sources.

McClintock, Canizares and Tarter (1975) list a series of C\,III, N\,III and O\,II lines
that should be detected in the case that non-selective emission is the predominant mechanism at work
(their Table 4).
The fact that we do not observe many of the strong lines
as C\,III 4186.9, 4325.6, 5695.9 and N\,III 4379.1, 4514.9, argues against this mechanism. The
selective emission of N\,III $\lambda \lambda$ 4634-4641 in the "Of star process" is also discarded since it should
produce absorptions at N\,III $\lambda \lambda$ 4097-4103, which are not observed.
Therefore our observations
favor the Bowen fluorescence mechanism for the $\lambda \lambda$ 4640-4650 emission in \ae, as
usual in symbiotic stars.
If this view is correct, then we should detect many OIII emission lines blueward 3450 \AA~
(McClintock, Canizares and Tarter, 1975).
AE\,Cir has not been hitherto observed in this region.

The emission line ratio N\,III 4641/4635 has been used as a diagnostic for electron density in hot plasmas produced in pulsed laser deposition experiments for thin film depositions (Bohigas, P\'erez-Tijerina \& Machorro 2004). These authors determined
population levels in a 10.000 K plasma considering radiative decay and collisional excitation and de-excitation. Their Fig.\,1 indicate that in the years 2000 and 2007, when N\,III 4641/4635 was between 0.7-1.1, the electron density was $\sim$ 10$^{10}$ or $\sim$ 10$^{17}$ cm$^{-3}$ (the case around 10$^{3}$ cm$^{-3}$ is excluded due to the absence of forbidden lines). On the other hand, in 2001 this ratio
increased to 2.5 (Fig.\,7 and Table 2). This value is out of the range predicted by
the aforementioned calculations, but an extrapolation of them implies an increase of electron
density between both epochs.

It is possible that the nebula in AE\,Cir is excited by X-rays, the existence of ionized helium
argues in favor of this view. AE\,Cir was covered during the ROSAT all-sky survey in August 1990
for about 550 sec (Aug 15-17 1990, HJD 2448118.9-2448120.9),
but no soft X-ray emission was detected. The object was not observed by XMM and Chandra,
and does not appear in the Galex UV survey. The non-detection by ROSAT implies an upper limit of 0.02
counts s$^{-1}$ in the PSPC-instrument. Assuming a blackbody spectrum with 50 (200) eV  and $E(B-V)$= 0.3, this implies an unabsorbed flux limit of 3 $\times 10^{-11} (5 \times 10^{-13}$) erg cm$^{-2}$ s$^{-1}$. Assuming a thermal bremsstrahlung spectrum with 1keV temperature gives 7.5 $\times 10^{-13}$ erg cm$^{-2}$ s$^{-1}$ and a power law with photon index 1.0, 1.5 or 2.0 gives $5, 6,$ or $8 \times
10^{-13}$ erg cm$^{-2}$ s$^{-1}$. This corresponds to a luminosity upper limit of 3.5 $\times 10^{33}$ -
6 $\times 10^{33}$ (D/10 kpc)$^{2}$ erg s$^{-1}$.
Note that if the distance were $\sim$ 10 kpc and the X-ray emission soft,
foreground absorption would make it particularly difficult to detect X-rays from AE\,Cir.

\subsection{On the He I emission and Balmer decrements}

The He\,I emission has been used as a diagnostic tool for the physical conditions in the
gaseous envelopes of symbiotic stars. We compared our He\,I emission line data with models by
Almog \& Netzer (1989) and Proga, Miko{\l}ajewska and Kenyon (1994), finding that they are in the upper
limit of symbiotic He\,I ratios, in the region of S-type symbiotics, and that they are not exactly reproduced by the models.
However, if we extrapolate the model results to AE\,Cir, the He\,I emission forming region would has
very high optical depth, electron density and temperature at the low state.
We note also that our He\,I ratios present significant deviations from Case B predictions, as usual in symbiotic stars.

We also compared our emission line strengths with predictions of the model for an illuminated red giant wind by Proga, Kenyon
\& Raymond (1998). These authors calculated equivalent widths and emission line ratios for the emergent spectrum of a red giant wind
illuminated by the hot component of a symbiotic binary system. Different wind velocity laws, mass loss rates and hot component temperature $T_{h}$ were used in the models. A comparison with the equivalent widths of AE\,Cir
shows that the H$\beta$ equivalent width is reproduced by a strong illumination model with $\log T_{h}$ decreasing 
from $\sim$ 4.8 in the low state to $\sim$ 4.6 in the high state, but the
He\,I 5875 and He\,II 4688 lines require $\log T_{h} \approx 5.2$. This could be interpreted in principle as two different forming regions for hydrogen and helium lines, but the predicted H$\beta$ equivalent width 
for $\log T_{h} \approx$ 5.2 is $\sim$ 100 \AA, something that is not observed.
In addition, we found that our He\,I emission line ratios are larger than expected. The failure of these models with the
He\,I $\lambda$6678/$\lambda$5875 emission line ratio is not restricted to AE\,Cir, but was noted to occur with symbiotic stars in general by Proga, Kenyon \& Raymond (1998).


Our Balmer decrements are listed in Table 3.
The general discrepancy with Case-B recombination theory probably indicate high densities and the importance of
atomic collisional excitation in the nebula. When comparing our decrements with those predicted by
Drake \& Ulrich (1980), we found that the 2007 data indicate an increase of the electron density at the high state.
This qualitative result is also obtained when comparing our Balmer decrements
with those predicted by the models for an
irradiated red giant atmosphere by Schwank, Schmutz and Nussbaumer (1997). The irradiated wind model of Proga, Kenyon
\& Raymond (1998), on the other hand, can reproduce the  H$\alpha$/H$\beta$ decrement observed in the high state only
with a model of high irradiation luminosity.

\begin{table}
\begin{minipage}{83mm}
\caption{Observed Balmer decrements compared with
predictions for Case-B recombination theory for different temperatures (Osterbrock \& Ferland 2006). Typical error for
AE\,Cir emission line ratios are 10\%.}
\begin{tabular}{lrccc}
\hline
Decrement &H$\alpha$/H$\beta$  &H$\gamma$/H$\beta$ &H$\delta$/H$\beta$ &H$\gamma$/HeII4686  \\
\hline
year-2000 & 4.25  & 0.36&   0.22&   0.31 \\
year-2001 & 5.01 & 0.51& 0.17& 0.25 \\
year-2007&2.36&0.40&0.37 &0.45 \\
2500 $K$&3.30 &0.44&0.24 & -  \\
5.000 $K$& 3.05 & 0.45 & 0.25& 0.56  \\
10.000 $K$&2.87&0.47&0.26&0.62\\
\hline
\end{tabular}
\end{minipage}
\end{table}

\section{General discussion}

As mentioned above, the presence of emission lines,
the Raman scattered line at $\lambda$ 6830 \AA~ and the detection of a cool stellar component
indicate a symbiotic classification for AE\,Cir. In a symbiotic system, a red giant
faces a hot component, normally a hot white dwarf, that ionizes part of the red giant wind.
Usually symbiotic star spectra reveal the red giant absorption lines, the emission lines coming from a
neutral region in the wind and also the high ionization species  formed
closer to the hot component. The rather early spectral type of $\sim$ K5, the hot 
ionizing source and the high nebular electron density indicate that AE\,Cir is
a new member of the small group of s-type yellow symbiotic stars (Schmid \& Nussbaumer 1993). 

The behavior in Fig.\,1 suggests the presence of recurrent outbursts in AE\,Cir. We discarded
total eclipses as explanation for the low states since traces of both components are observed 
in the high and low states.
However, the outbursts are not reminiscent of typical symbiotic novae
outbursts. While symbiotic novae have outburst amplitudes of a few magnitudes,
the rise time is of the order of years, and the decay time of order of
decades. In the case of AE Cir, the rise time in 1989
(HJD $\sim$244\,7700) could be only 2 weeks (if we trust 
in the two visual observations discussed previously)
and the next rise happened
already in 2001, 12 years after the previous one. Compared to all other
known symbiotic novae, AE\,Cir would be on the very short-timescale
end. The outbursts observed in AE\,Cir can be better
classified as of the type called "classical symbiotic outburst" (Sokoloski 2003).
During a classical symbiotic outburst, the optical brightness typically increases by several
magnitudes, but in some cases can increase by as little as one magnitude. The system
may take weeks or months to reach maximum brightness, and then months or years
to fade. However a peculiar feature of the AE\,Cir light curve is the long time spend at
maximum: the low state has $\sim$ 38\% the length of the high state.
The general cause for classical symbiotic outbursts has not yet been established (Sokoloski 2003), although
in Z And one outburst was identified starting with
characteristics of a dwarf nova outburst (i.e. triggered by a disc instability) and
followed by a nova-type outburst (i.e. probably caused by enhanced nuclear burning on the surface of the white dwarf,
Sokoloski et al. 2006). Incidentally, the shape of the main outburst of \ae~ is remarkably similar to a
superoutburst of a SU UMa type dwarf nova, which is explained by a thermal-tidal disc instability (e.g. Warner 1995).

We have calculated the temperature of the hot component using the
relation of Murset \& Nussbaumer (1994) that uses the ionization potential of the highest-ionization species
visible in the spectrum $\chi_{max}$. This relation is $T_{h}$ = 1000 $\chi_{max}$ and
gives 100\,000 $K$ in the low state in 2000 and 2001 and 55\,000 $K$ in the high state in 2007.
We also used the Iijima (1981) relationship for a radiation-bounded nebula, based on the ratio of
nebular emission lines. We obtained upper limits for $T_{h}$ of 208\,000 $K$, 268\,000 $K$ and 242\,000
$K$ in 2000, 2001 and 2007. The lower calculated temperature of the hot component in the high state is deduced from the disappearance
of the 6830 \AA~ emission feature. We observe that the decrease of $T_{h}$ argues against the accretion disc instability
model for the outbursts of AE\,Cir. In this model the disc increases luminosity during outburst due to an increase of mass accretion rate
(e.g. Duschl 1986),
but in this case we should observe an increase of the hot component temperature.
On the other hand, the nuclear burning model implies an expansion of the white dwarf photosphere at constant bolometric luminosity and predicts a decrease of the hot component temperature, as observed. However, the very fast development of the main outburst is contrary to the model predictions; 
the shell envelope  should expand on the thermal time scale,
which is years to decades for typical conditions (Iben 1982, Iben \& Tutukov 1996). 
The second outburst instead has a much more gradual development, 
more in agreement with the nuclear burning model. However, 
we do not observe spectral features 
that could reveal an expanded white dwarf atmosphere, only the presence of
a blue continuum at the high state (Fig.\,5). In particular, no 
A-type or F-type shell spectrum is observed as in the 
symbiotic stars CI Cyg and RR Tel at the high state (Belyakina 1979, Thackeray 1950).
A K-band spectrum of AE\,Cir kindly taken by Elisa Nespoli (private communication)
with the ESO NTT SOFI in May 25, 2007 (MJD= 2454245.9942) with a S/N of 35
shows no Br$\gamma$ emission or absorption, indicating that the nebular lines 
are not prominent in the infrared at the high state, and that there are no signs of 
an hypothetical hot stellar component. The red giant is not detected in this spectrum,
possibly due to the low S/N of the data.  

If full visibility of the red giant occurs at the low state at $V \sim 15.5$,
this would indicate a distance of 9.4 kpc, for
a typical K5 giant behind the galactic disc. This should place
\ae~  below 1.4 kpc from the galactic plane.
We also note that if the high state is due to the
hot component outburst, its $V \sim$ 13 mag combined with $\log T_{h} \ga$ 4.6 should imply
a very high $M_{bol}  \ga$ -6.4,  above the Eddington limit for a typical 0.6 M$_{\odot}$ 
symbiotic white dwarf, but around the Eddington limit for a massive white dwarf accretor of
$\sim$ 1 M$_{\odot}$ (Belczy\'nski et al. 2000, Miko{\l}ajewska 2003).

In summary, while the behavior of $T_{h}$ and the long outburst rising time 
could  suggest nuclear burning, the blue continuum and
the lack of A-type or F-type shell spectrum at the high state 
point to the accretion disc instability model.  We now examine 
if the explanation for the outbursts of AE\,Cir
could be related to the bizarre behavior of the spectral features of the secondary star.


It is remarkable that we 
observe the TiO band around $\lambda$ 6300 \AA~ at 
roughly comparable depth in all spectra of Fig.\,5. 
The difference in $V$ magnitude between the
2007 and 2000 spectra is about 1.5 magnitudes. This should imply a change of the relative 
contribution of the cool giant to the total light by a factor 4, 
but the change in equivalent width of the TiO band is only 25\%. 
In other words, it should be practically impossible to detect the
secondary star at maximum if its brightness remains the same that at minimum.
The fact that we observe the secondary star contributing roughly the same fraction of light during the outburst
cycle  suggests that: (i) the star varies its brightness or (ii) the whole system light is obscured by some 
source during the low state. At the same time, the spectral type determinations 
indicate that color and temperature of the giant star remained roughly unchanged during the observation epochs.

From Fig.\,1 the overall brightness at the epoch of the SMARTS observation is $V \approx$ 12.7 mag. 
If we assume that half of the luminosity comes from the giant star we get $V \sim 13.5$ for the giant star.
Using this value and $E(B-V)$ = 0.26 we obtain a distance of 10.2 kpc, in nice agreement with our earlier estimate.
From the above it seems that the putative change of the stellar brightness of the giant star in AE\,Cir could be large. 
A $\sim$ 2 mag change at constant radius should imply a change in temperature by a factor of 1.6, which is not observed.  This means that the giant star in AE\,Cir should change its radius by a factor $\sim$ 2, likely engulfing their companion. 
Such a big change in radius is hard to understand.

Regarding the alternative of obscuration, it is compatible with the larger reddening observed at minimum, as derived from 
the strength of the NaD doublet and system color, but not with the larger emission observed at minimum neither with the long decrease of brightness observed after maximum. We conclude that the 
data present difficulties for the canonical symbiotic outburst models.

We note that if the ellipsoidal like variability with a periodicity of 342 days 
is the binary period, the amplitude, 
if ellipsoidal in nature, largely exceeds the expected values for a tidally distorted star
(Soszynski et al. 2004, Morris 1985).  We note that the $B-R$ color shows a systematic change over the 342 day
period (see Fig.\,3), being maximum at the time of minimum
flux of the 342 day oscillation. This would be expected if
the cool star is in front of the hot component.
If this were true, then the rather narrow eclipse light curve observed in Fig.\,3
could indicate that the red giant self-eclipses a zone of chromospheric
fluorescence, as proposed for V1329 Cyg (Chochol \& Wilson 2001). The visibility 
of the cool giant at minimum and maximum could in principle allow detection of such a variability.
If we really observe a chromospheric eclipse in the SMARTS data,
the 40-day eclipse length suggests a radius for the red giant of:\\

$R_{g} = \frac{1}{f} \times \frac{40}{342} \times 2\pi a = \frac{0.74}{f}a $\hfill(2) \\

\noindent
where $a$ is the binary separation and $f$ a geometric factor depending on the characteristics of the
illuminated surface. If $f$ = 2, i.e. the eclipse corresponding length is comparable to the stellar diameter,
the radius is surprisingly close to the Roche-lobe radius for the secondary star in a binary of mass ratio unity,
viz.\, 0.38$a$ (Warner 1995). As the mass of a K5 giant is $\sim$ 1.2 M$_{\odot}$ and the mass of a white dwarf
(assumed to be the hot component)
is around 1 M$_{\odot}$, then it is possible  that the red giant in AE\,Cir fills its Roche-lobe.
If we use the above masses for the components and the third Kepler law we obtain $a$  = 1.87 $\times$ 10$^{13}$ cm and
$R_{g}$ = 6.92 $\times$ 10$^{12}$ cm = 99 R$_{\odot}$.
The condition of Roche-lobe overfilling, along with the putative orbital period of 342 days,
constrains the stellar mean density to $\log{\overline{\rho}}$ = -5.8 g cm$^{-3}$ (e.g. Warner 1995).
This value yield an independent estimate for the red giant mass of 1.1 M$_{\odot}$, in nice agreement with our early assumption.
We note that the derived radius is in the upper limit of radius for red giants with a temperature of 3950 $K$ (van Belle et al. 1999). Models for irradiated late type giants show an expansion of the atmosphere by several percents for large
hot component luminosities (Proga et al. 1996), hence irradiation could play a role in the size of the red giant in AE\,Cir. Irradiation is also suggested by the emission line ratios discussed in the previous section. 
The above calculation implies a luminosity for the red star of $L_{bol}$ = 7.80 $\times$ 10$^{36}$ erg s$^{-1}$, yielding $M_{bol}$ = -3.42 and using a bolometric correction of $BC$ = -1.02 for a K5 giant suggests $M_{V}$ = -2.4.

\section{Conclusions}

In this paper we have investigated the nature of the symbiotic star candidate AE\,Cir. We have analyzed
new optical photometric and spectroscopic data, 2MASS infrared photometry and 24 years of  visual photometry in an integrated
way in order to get a better understanding of the system. Our conclusions can be summarized as follows:\\

\begin{itemize}

\item The symbiotic nature is confirmed.
This result is based on the detection of the $\lambda$ 6830 \AA~ emission
feature and the spectral signatures of a cool stellar component of spectral type $\sim$ K5.

\item The spectral type $\sim$ K5 and the spectral and photometric properties 
indicate that \ae~ is a member of the small group of yellow symbiotic stars of the s-type.

\item The light curve is characterized by outbursts lasting $\sim$ 4000 days and overall
amplitude of variability about 4 magnitudes. The outbursts show a slow decline of
$\sim 2 \times 10^{-4}$ mag/day. 
The duration  of the low state is about 38\% the high state.

\item A strong signal at 342 $\pm$ 15 days is detected in the light curve. The light curve
folded with this period shows two broad minima with different amplitude. 

\item The spectrum in the low state is characteristic of a symbiotic star without forbidden lines and
very strong He\,I\,4868 emission.
At maximum the emission feature at $\lambda$ 6380 \AA~ disappears, the spectrum becomes flat and the relative
intensity of the He\,I 4686 \AA~line, compared with the Balmer lines, becomes lower.  At the same time 
the overall emission line spectrum shows smaller equivalent widths.
Signatures of the cool companion are observed at the high and low state. 

\item We observe H\,I, He\,I and He\,II emission line variability in time scales
of minutes in 2001 being larger in the line centers. The Na\,D and $\lambda$ 6820 \AA~ lines do not follow this variability.
We also observe larger broadening and asymmetry in the Balmer lines compared with
others emission lines in this epoch.

\item The data suggest that full visibility of the red giant should occur at the low state at $V \sim 15.5$. For
a typical K5 giant behind the galactic disc this would indicate a distance of 9.4 kpc. This should indicate that
\ae~  is about 1.4 kpc below the galactic plane. 

\item We suggest that an eclipse-like event could be interpreted as self-occultation of the irradiated red giant hemisphere.
In this case we found that it is possible that the red giant fills its Roche-lobe. Assuming a white dwarf
hot component of 1 M$_{\odot}$, we estimate a radius of 99 R$_{\odot}$, a mass of 1.1 M$_{\odot}$ and
M$_{V}$ = -2.4 for the red giant at the epoch of the SMARTS observations.

\item The atypical character of \ae~ is revealed in 
the long time passing in outburst and the dwarf nova superoutburst shape of the light curve. 
Another atypical feature is the rather strong 
TiO absorption band around $\lambda$ 6300 \AA~ at the high state.  This could indicate that the
red giant and the hot component are partially obscured at minimum 
or the brightness of the secondary star
is not constant. Obscuration hardly explains the large line emission observed at minimum and the long brightness
decay after maximum whereas variations of the red giant brightness imply unrealistic changes in the stellar radius.

\item  At present, the available data for AE\,Cir are hard to reconciliate with canonical models for symbiotic outbursts.


\end{itemize}

\section*{Acknowledgments}
 
We acknowledge an anonymous referee for her/his help in improving
the first version of this manuscript. We acknowledge with thanks the variable star observations from the AAVSO International
Database contributed by observers worldwide and used in this research.
This research has made use of the AFOEV database
operated at CDS, France. We
appreciate the enormous amount of observing time by the
amateurs contributing to the AAVSO and AFOEV light curve.
We are grateful to Lex Kaper and Arjen van der Meer
(Amsterdam) for the observations with the Dutch 0.9m telescope. We are grateful to Dr. A. K\"upc\"u Yolda\c{s}
for help in the analysis of the SMARTS data set. We acknowledge Elisa Nespoli and Juan Fabregat for
kindly take a K-band spectrum of AE\,Cir on May 2007. REM acknowledges support by Grants
Fondecyt \#1030707 and \#1070705.

\bsp

\label{lastpage}

\end{document}